1# Microscopic theory of the atom work function from the binary mixture of simple liquids. 1. General results

S.S. Kaim

*National University of Odessa, Odessa, Ukraine*

kaimss@inbox.ruAnalyzing the equations for the unary distribution functions of the Bogolubov-Born-Green-Kirkwood-Yvon chain of equations for the equilibrium two-phase system consisting of binary mixture of simple liquids and gas mixture with plane interface, we derived the analytical expression for the mono-atomic potentials in liquid and gas phases. Using the asymptotic values of mono-atomic potentials in liquid and in gas far away from interface, for each component of the mixture was obtained expression for the atom work function from liquid into gas. The interrelation between the general equation of state for liquid mixture and the atom work functions from mixture into vacuum was established. The stability criterion for each component of the mixture in limiting points of the first type (using I.Z. Fisher's classification) was formulated in terms of the atom work function from liquid mixture into vacuum. As it turned out, the stability criteria correspond to atomization conditions of the mixture components.## 1. INTRODUCTION

Information about thermodynamic and phase equilibrium properties of multi-component mixtures has a crucial importance in many industrial processes (petroleum refining, gas mixtures separation, obtaining gases from air, supercritical extraction of substances, etc.). At the microscopic level the phase behavior of liquid and gas mixtures is determined by the activation processes that play an important role both on equilibrium and non-equilibrium conditions. Activation processes may hold mono- and poly- atomic nature, and are controlled with the effective self-consistent mono-atomic fields created by all atoms of the system (pair and non-additive inter-atomic interactions and correlations). At extreme and non-equilibrium thermodynamic conditions in combination with nano-liter volumes of mixture (micro-fluidics), the specific behavior of the self-consistent potentials and their strong influence on the macroscopic response of the mixture, will probably lead to a novel effects that require a separate consideration. Thus, the effective mono-atomic potentials define the behavior of inhomogeneous mixture and at the same time strongly depend on the external thermodynamic conditions. As a rule, the effective mono-atomic potentials have a different spatial dependence for the each component of the mixture. Therefore, for example, coexisting liquid and gas phases have different concentrations of the components. The mono-atomic potentials define the conditions of the relative stability of mixture, conditions of the transition to the metastable or absolutely unstable states. One of the important thermodynamic characteristics directly connected with the level of the self-consistent mono-atomic potential, is the atom work function from the liquid phase into the gas. The atom work function characterizes the absolute stability of matter. It is defined [1] as a difference in levels of the total self-consistent mono-atomic potential that act on a separate atom deep inside and far outside the matter. In other words it is a work that needs to be done to move a separate atom from a point inside the matter (the point should be far enough from the interface) into vacuum and leave it there with a velocity equal to zero.

The microscopic theory of the atom work function from the simple liquids was developed in [1]. Asymptotes of the self-consistent mono-atomic potential were obtained from the first equation of the Bogolubov-Born-Green-Kirkwood-Yvon (BBGKY) chain of equations for the atom unary distribution function. The calculations of the work function were performed for the temperature interval from melting point up to the critical temperature, and in spite of high sensitivity of the results to choice of the atom pair distribution function, gave a good agreement with the experimental measurements of evaporation heat.

The work [2] performed the logical extension to the work function theory [1] and established the coupling between the atom work function from liquid into vacuum and the stability criterion of liquid in limiting points of the first type (using I.Z. Fisher's classification [3,4]). The obtained temperature and volumetric data for the atom work function from liquid into vacuum clearly show a possibility to reach states of the liquid with a negative atom work function that corresponds to absolutely unstable states of the liquid with respect to atomization. The assumption that these states may appear in liquid during a collapse of cavity bubbles gives a possible explanation of the observed experimentally sonoluminescence effect. The authors [2] developed the corresponding two-stage emission-impact mechanism of sonoluminescence based on non-thermal processes in liquid in a state with a negative atom work function. However, the experimental observations of sonoluminescence are mostly performed in binary mixtures (mixtures of water, sulphuric acid, acetone and rare gases of low concentrations), and showed high sensitivity of sonoluminescence intensity to concentrations and temperature intervals [5]. Discovered in experiments essential concentration and temperature dependences of sonoluminescence efficiency in liquids [5] indicate molecular-kinetic origins of this phenomenon. Study of the phenomenon of atoms emission from surfaces of liquid mixtures as a possible physical basis of sonoluminescence is of the great interest. The atom work function is a crucial physical quantity required for calculation of atoms emission from surfaces of liquid mixtures. For elucidation of absolute stability limits with respect to each component of the mixture it is necessary to find general thermodynamic conditions of possible spontaneous atoms emission from liquid mixtures. The aim of this work is a study of the atom work function from a simple liquid mixture to a coexisting gas mixture and a study of general conditions of absolute stability of liquid mixtures and gas mixtures with respect to atomization of their components. The analysis of the problem is performed in terms of most general conditions of thermodynamic and mechanical equilibrium of two-phase systems "mixture of simple liquids – mixture of gases". Such conditions of equilibrium for two coexisting phases can be expressed using the BBGKY chain of equations for particle group distribution functions in liquid and gas phases. The atom work functions for every component from the liquid mixture to the gas mixture can be expressed in terms of asymptotes of monatomic potentials for each component in the liquid and in the gas phases. Expressions for effective monatomic potentials were obtained by analyzing the equations for the unary distribution functions of the BBGKY chain of equations. In the work there is also made a generalization of the I.Z.Fisher's stability criterion for simple liquids in limiting points of the first type [1-4] for case of a simple liquid mixture and a gas mixture. It was shown that the obtained stability criteria for each component of mixtures correspond to atomization conditions of components.

## 2. ATOM WORK FUNCTION FROM LIQUID MIXTURES

We consider the equilibrium two-phase two-component system "liquid mixture – gas mixture" with a flat interface, atoms of which interact by means of central pair forces. The study of properties of such system will be carried out in a framework of the Gibbs canonical ensemble. The classical Hamiltonian of such system can be written as

$$\hat{H} = \sum_{\alpha=1}^{2} \sum_{i=1}^{N_\alpha} \frac{\mathbf{P}_i^{(\alpha)2}}{2M_\alpha} + \frac{1}{2} \sum_{\alpha=1}^{2} \sum_{\beta=1}^{2} \sum_{i=1}^{N_\alpha} \sum_{j=1}^{N_\beta} \Phi_{\alpha\beta}\left(\mathbf{R}_i^{(\alpha)} - \mathbf{R}_j^{(\beta)}\right),  \qquad (1)$$

where $\mathbf{P}_i^{(\alpha)}, M_\alpha$ are the momentum and the mass of the atom of the type $\alpha$; $N_\alpha$ is the number of atoms of the type $\alpha$; $\Phi_{\alpha\beta}\left(\mathbf{R}_i^{(\alpha)} - \mathbf{R}_j^{(\beta)}\right)$ is the interaction potential energy of two atoms of types $\alpha$ and $\beta$. The statistical sum of the system can be written as [6-8]

$$Z_{N_1,N_2} = \frac{1}{N_1! \cdot N_2!} \left(\frac{M_1 k_B T}{2\pi\hbar^2}\right)^{\frac{3}{2}N_1} \left(\frac{M_2 k_B T}{2\pi\hbar^2}\right)^{\frac{3}{2}N_2} Q_{N_1,N_2}, \qquad (2)$$

where $Q_{N_1,N_2}$ is the configuration integral of the system, which is expressed as

$$Q_{N_1,N_2} = \int \ldots \int d^3 R_1^{(1)} \ldots d^3 R_{N_2}^{(2)} \exp\left[-\frac{V\left(\mathbf{R}_1^{(1)},\ldots,\mathbf{R}_{N_2}^{(2)}\right)}{k_B T}\right], \qquad (3)$$

where $V\left(\mathbf{R}_1^{(1)},\ldots,\mathbf{R}_{N_2}^{(2)}\right)$ is the potential energy of the system of interactive atoms of the two-component two-phase system.

Description of the system can be carried out in terms of the particle group distribution functions [6-8]. The Gibbs distribution function for a system consisting of $N_1$ particles of the first type and of $N_2$ particles of the second type can be written as

$$D_{N_1,N_2}\left(\mathbf{R}_1^{(1)},...,\mathbf{R}_{N_1}^{(1)},\mathbf{R}_1^{(2)},...,\mathbf{R}_{N_2}^{(2)}\right) = \frac{1}{Q_{N_1,N_2}} \exp\left(-\frac{V\left(\mathbf{R}_1^{(1)},...,\mathbf{R}_{N_2}^{(2)}\right)}{k_B T}\right). \tag{4}$$

For the binary inhomogeneous mixture with interface between liquid and gas, we assume the following definition of the unary distribution functions [8]

$$F_1^{(\alpha)}\left(\mathbf{R}^{(\alpha)}\right) = N_\alpha \int ... \int \delta\left(\mathbf{R}^{(\alpha)} - \mathbf{R}_s^{(\alpha)}\right) D_{N_1,N_2}\left(\mathbf{R}_1^{(1)},...,\mathbf{R}_{N_2}^{(2)}\right) d^3 R_1^{(1)}...d^3 R_{N_1}^{(1)} d^3 R_1^{(2)}...d^3 R_{N_2}^{(2)}, \alpha = 1,2, \tag{5}$$

where $\delta\left(\mathbf{R}^{(\alpha)} - \mathbf{R}_s^{(\alpha)}\right)$ is the Dirac delta function. Normalization requirements for the unary distribution functions defined above can be written as

$$\int d^3 R^{(\alpha)} \cdot F_1^{(\alpha)}\left(\mathbf{R}^{(\alpha)}\right) = N_\alpha, \quad \alpha = 1,2. \tag{6}$$

The unary distribution functions defined like this describe a local density of a number of atoms of each type in the binary inhomogeneous mixture including the layer near the two phases interface. The atom pair distribution functions in the binary inhomogeneous mixture we define correspondingly as [8]

$$F_2^{(\alpha\beta)}\left(\mathbf{R}^{(\alpha)},\mathbf{R}^{(\beta)}\right) = N_\alpha (N_\beta - \delta_{\alpha\beta}) \int ... \int \delta\left(\mathbf{R}^{(\alpha)} - \mathbf{R}_s^{(\alpha)}\right) \delta\left(\mathbf{R}^{(\beta)} - \mathbf{R}_j^{(\beta)}\right) \times$$
$$\times D_{N_1,N_2}\left(\mathbf{R}_1^{(1)},...,\mathbf{R}_{N_2}^{(2)}\right) d^3 R_1^{(1)}...d^3 R_{N_1}^{(1)} d^3 R_1^{(2)}...d^3 R_{N_2}^{(2)}, \alpha,\beta = 1,2. \tag{7}$$

Using standard methods for obtaining the BBGKY chain of equations and using the definitions (5)-(7) we obtain the equations for the unary atom distribution functions in the inhomogeneous two-component mixture "liquid – gas" [8]

$$\nabla_1^{(1)} F_1^{(1)}\left(\mathbf{R}_1^{(1)}\right) = -\frac{1}{k_B T} \int d^3 R_2^{(1)} \cdot \nabla_1^{(1)} \Phi_{11}\left(\mathbf{R}_1^{(1)} - \mathbf{R}_2^{(1)}\right) F_2^{(11)}\left(\mathbf{R}_1^{(1)},\mathbf{R}_2^{(1)}\right) -$$
$$-\frac{1}{k_B T} \int d^3 R_1^{(2)} \cdot \nabla_1^{(1)} \Phi_{12}\left(\mathbf{R}_1^{(1)} - \mathbf{R}_1^{(2)}\right) \cdot F_2^{(12)}\left(\mathbf{R}_1^{(1)},\mathbf{R}_1^{(2)}\right), \tag{8}$$

$$\nabla_1^{(2)} F_1^{(2)}\left(\mathbf{R}_1^{(2)}\right) = -\frac{1}{k_B T} \int d^3 R_2^{(2)} \cdot \nabla_1^{(2)} \Phi_{22}\left(\mathbf{R}_1^{(2)} - \mathbf{R}_2^{(2)}\right) F_2^{(22)}\left(\mathbf{R}_1^{(2)},\mathbf{R}_2^{(2)}\right) -$$
$$-\frac{1}{k_B T} \int d^3 R_1^{(1)} \cdot \nabla_1^{(2)} \Phi_{21}\left(\mathbf{R}_1^{(2)} - \mathbf{R}_1^{(1)}\right) \cdot F_2^{(21)}\left(\mathbf{R}_1^{(2)},\mathbf{R}_1^{(1)}\right). \tag{9}$$

For analysis of the self-consistent monatomic potentials we use the following representations for the pair atom distribution functions

$$F_2^{(\alpha\beta)}\left(\mathbf{R}_1^{(\alpha)},\mathbf{R}_2^{(\beta)}\right) = F_1^{(\alpha)}\left(\mathbf{R}_1^{(\alpha)}\right) \cdot F_1^{(\beta)}\left(\mathbf{R}_2^{(\beta)}\right) \cdot g^{(\alpha\beta)}\left(\mathbf{R}_1^{(\alpha)},\mathbf{R}_2^{(\beta)}\right), \alpha,\beta = 1,2, \tag{10}$$

where new functions $g^{(\alpha\beta)}\left(\mathbf{R}_1^{(\alpha)},\mathbf{R}_2^{(\beta)}\right)$ were introduced, and they describe pair correlations in the inhomogeneous mixture (both in the liquid and in the gas phases). Hereinafter for simplification of expressions the type of atoms in the radius vectors of the particular atoms will be omitted. Taking into account the representation (10) the equations (8) and (9) can be written in the following form

$$\nabla_1 \ln F_1^{(1)}(\mathbf{R}_1) = -\frac{1}{k_B T} \int d^3 R_2 \cdot F_1^{(1)}(\mathbf{R}_2) \cdot \nabla_1 \Phi_{11}(\mathbf{R}_1 - \mathbf{R}_2) \cdot g^{(11)}(\mathbf{R}_1, \mathbf{R}_2) -$$

$$-\frac{1}{k_B T} \int d^3 R_2 \cdot F_1^{(2)}(\mathbf{R}_2) \cdot \nabla_1 \Phi_{12}(\mathbf{R}_1 - \mathbf{R}_2) \cdot g^{(12)}(\mathbf{R}_1, \mathbf{R}_2), \tag{11}$$

$$\nabla_1 \ln F_1^{(2)}(\mathbf{R}_1) = -\frac{1}{k_B T} \int d^3 R_2 \cdot F_1^{(2)}(\mathbf{R}_2) \cdot \nabla_1 \Phi_{22}(\mathbf{R}_1 - \mathbf{R}_2) \cdot g^{(22)}(\mathbf{R}_1, \mathbf{R}_2) -$$

$$-\frac{1}{k_B T} \int d^3 R_2 \cdot F_1^{(1)}(\mathbf{R}_2) \cdot \nabla_1 \Phi_{21}(\mathbf{R}_1 - \mathbf{R}_2) \cdot g^{(21)}(\mathbf{R}_1, \mathbf{R}_2). \tag{12}$$

We can use the fact that in case of a flat interface between two phases, the equation of which is chosen as $z = 0$, the unary distribution functions depend on one coordinate $z$. Far from the interface in the liquid mixture ($z \to -\infty$) the asymptotes of the unary distribution functions are equal to the corresponding atoms number density for volume

$$\lim_{z \to -\infty} F_1^{(1)}(z) = c_1 n, \quad \lim_{z \to -\infty} F_1^{(2)}(z) = c_2 n, \tag{13}$$

where $c_1, c_2$ are the concentrations of the mixture components in the liquid phase; $n$ - the atoms number density in the liquid phase. Similarly, far from the interface in the gas mixture ($z \to \infty$)

$$\lim_{z \to \infty} F_1^{(1)}(z) = c_1' n', \quad \lim_{z \to \infty} F_1^{(2)}(z) = c_2' n', \tag{14}$$

where $c_1', c_2'$ are the concentrations of the mixture components in the gas phase; $n'$ - the atoms number density in the gas phase.

Taking into account (13) the equations (11) and (12) can be written in the equivalent integrated forms

$$\begin{cases} \dfrac{F_1^{(1)}(z)}{c_1 n} = \exp\left\{-\dfrac{1}{k_B T}\left[\int\limits_{-\infty}^{z} dz_1 \int d^3 R_2 \cdot F_1^{(1)}(z_2) \cdot \dfrac{d\Phi_{11}(\mathbf{R}_1 - \mathbf{R}_2)}{dz_1} \cdot g^{(11)}(\mathbf{R}_1, \mathbf{R}_2) + \right.\right. \\ \left.\left. + \int\limits_{-\infty}^{z} dz_1 \int d^3 R_2 \cdot F_1^{(2)}(z_2) \cdot \dfrac{d\Phi_{12}(\mathbf{R}_1 - \mathbf{R}_2)}{dz_1} \cdot g^{(12)}(\mathbf{R}_1, \mathbf{R}_2)\right]\right\}, \\ \dfrac{F_1^{(2)}(z)}{c_2 n} = \exp\left\{-\dfrac{1}{k_B T}\left[\int\limits_{-\infty}^{z} dz_1 \int d^3 R_2 \cdot F_1^{(2)}(z_2) \cdot \dfrac{d\Phi_{22}(\mathbf{R}_1 - \mathbf{R}_2)}{dz_1} \cdot g^{(22)}(\mathbf{R}_1, \mathbf{R}_2) + \right.\right. \\ \left.\left. + \int\limits_{-\infty}^{z} dz_1 \int d^3 R_2 \cdot F_1^{(1)}(z_2) \cdot \dfrac{d\Phi_{21}(\mathbf{R}_1 - \mathbf{R}_2)}{dz_1} \cdot g^{(21)}(\mathbf{R}_1, \mathbf{R}_2)\right]\right\}. \end{cases} \tag{15}$$

The system of the equations (15) is self-consistent with respect to the unary distribution functions. If the functions in the indexes of powers are known then the expressions in the square brackets can be considered as the self-consistent single-particle potentials which affect particular atoms of each component, both in the liquid and in the gas phases. At the same time non-local dependences of the partial functions $g^{(\alpha\beta)}(\mathbf{R}_1, \mathbf{R}_2)$ certainly have to be taken into account. Such monatomic potentials can be written as

$$U_1(z) = \int_{-\infty}^{z} dz_1 \int d^3 R_2 \cdot F_1^{(1)}(z_2) \cdot \frac{d\Phi_{11}(\mathbf{R}_1 - \mathbf{R}_2)}{dz_1} \cdot g^{(11)}(z_1, z_2, \rho_{12}^{\parallel}) +$$

$$+ \int_{-\infty}^{z} dz_1 \int d^3 R_2 \cdot F_1^{(2)}(z_2) \cdot \frac{d\Phi_{12}(\mathbf{R}_1 - \mathbf{R}_2)}{dz_1} \cdot g^{(12)}(z_1, z_2, \rho_{12}^{\parallel}), \qquad (16)$$

$$U_2(z) = \int_{-\infty}^{z} dz_1 \int d^3 R_2 \cdot F_1^{(2)}(z_2) \cdot \frac{d\Phi_{22}(\mathbf{R}_1 - \mathbf{R}_2)}{dz_1} \cdot g^{(22)}(z_1, z_2, \rho_{12}^{\parallel}) +$$

$$+ \int_{-\infty}^{z} dz_1 \int d^3 R_2 \cdot F_1^{(1)}(z_2) \cdot \frac{d\Phi_{21}(\mathbf{R}_1 - \mathbf{R}_2)}{dz_1} \cdot g^{(21)}(z_1, z_2, \rho_{12}^{\parallel}). \qquad (17)$$

In the formulas (16) and (17) coordinate dependences of the correlation functions for the flat geometry of the problem were taken into account evidently.

Inasmuch as the problem of search of the pair correlation functions is not solved even for an one-component system "liquid – gas", then for calculations of the monatomic potential asymptotes far from the interface in the liquid and in the gas phases we use the following approximation

$$g^{(\alpha\beta)}(\mathbf{R}_1^{(1)}, \mathbf{R}_2^{(1)}) \approx \overline{F}_2^{(\alpha\beta)}(|\mathbf{R}_1^{(\alpha)} - \mathbf{R}_2^{(\beta)}|), \qquad (18)$$

where $\overline{F}_2^{(\alpha\beta)}(|\mathbf{R}_1^{(\alpha)} - \mathbf{R}_2^{(\beta)}|)$ are the partial pair atom distribution functions in the homogeneous liquid mixture or in the gas mixture which correspond to concentrations of the components far from the interface. For calculation of the asymptotes of the monatomic potentials for the unary distribution functions in the liquid phase we use the approximation

$$F_1^{(1)}(z) \approx c_1 n \Theta(-z), \quad F_1^{(2)}(z) \approx c_2 n \Theta(-z). \qquad (19)$$

The calculation of the asymptotes of the monatomic potentials using the approximation (19) corresponds to the monatomic potentials deep in the liquid and far from the surface of the liquid in vacuum. Taking into account the approximations (18) and (19) the expressions for the monatomic potentials can be written as

$$U_1(z) = -2\pi n c_1 \int_0^{\infty} dR \frac{d\Phi_{11}(R)}{dR} \cdot \overline{F}_2^{(11)}(R) \cdot f(z, R) - 2\pi n c_2 \int_0^{\infty} dR \cdot \frac{d\Phi_{21}(R)}{dR} \cdot \overline{F}_2^{(12)}(R) \cdot f(z, R), \qquad (20)$$

$$U_2(z) = -2\pi n c_2 \int_0^{\infty} dR \frac{d\Phi_{22}(R)}{dR} \cdot \overline{F}_2^{(22)}(R) \cdot f(z, R) - 2\pi n c_1 \int_0^{\infty} dR \cdot \frac{d\Phi_{21}(R)}{dR} \cdot \overline{F}_2^{(21)}(R) \cdot f(z, R), \qquad (21)$$

where the function $f(z, R)$ is the following designation

$$f(z, R) \equiv \Theta(z - R) \cdot (-2/3) \cdot R^3 + \Theta(R - z) \cdot \Theta(z + R) \cdot [z^3/6 - R^3/3 + Rz^2 - R^2 z/2].$$

The asymptotes of the monatomic potentials (20) and (21) can be easily found. Thus far from the surface of the liquid in vacuum we obtain the expressions

$$\lim_{z \to +\infty} U_1(z) = \frac{4\pi}{3} nc_1 \int_0^\infty dR \frac{d\Phi_{11}(R)}{dR} \cdot \overline{F}_2^{(11)}(R) \cdot R^3 + \frac{4\pi}{3} nc_2 \int_0^\infty dR \cdot \frac{d\Phi_{21}(R)}{dR} \cdot \overline{F}_2^{(12)}(R) \cdot R^3, \tag{22}$$

$$\lim_{z \to +\infty} U_2(z) = \frac{4\pi}{3} nc_2 \int_0^\infty dR \frac{d\Phi_{22}(R)}{dR} \cdot \overline{F}_2^{(22)}(R) \cdot R^3 + \frac{4\pi}{3} nc_1 \int_0^\infty dR \cdot \frac{d\Phi_{21}(R)}{dR} \cdot \overline{F}_2^{(21)}(R) \cdot R^3. \tag{23}$$

Far from the surface in the mixture of liquids we obtain

$$\lim_{z \to -\infty} U_1(z) = 0, \quad \lim_{z \to -\infty} U_2(z) = 0. \tag{24}$$

Similarly, for the monatomic potentials created by the semi-bounded gas mixture for which the unary functions are assigned as

$$F_1^{(1)}(z) \approx c_1' n' \Theta(z), \quad F_1^{(2)}(z) \approx c_2' n' \Theta(z), \tag{25}$$

we obtain the expressions

$$U_1'(z) = -2\pi n' c_1' \int_0^\infty dR \frac{d\Phi_{11}(R)}{dR} \cdot \overline{F}_2'^{(11)}(R) \cdot f(z, R) - 2\pi n' c_2' \int_0^\infty dR \cdot \frac{d\Phi_{21}(R)}{dR} \cdot \overline{F}_2'^{(12)}(R) \cdot f(z, R), \tag{26}$$

$$U_2'(z) = -2\pi n' c_2' \int_0^\infty dR \frac{d\Phi_{22}(R)}{dR} \cdot \overline{F}_2'^{(22)}(R) \cdot f(z, R) - 2\pi n' c_1' \int_0^\infty dR \cdot \frac{d\Phi_{21}(R)}{dR} \cdot \overline{F}_2'^{(21)}(R) \cdot f(z, R), \tag{27}$$

where $\overline{F}_2'^{(\alpha\beta)}(R)$ are the partial pair distribution functions of atoms in the homogeneous gas mixture.

For the asymptotes of the monatomic potentials (26) and (27) we obtain

$$\lim_{z \to -\infty} U_1'(z) = \frac{4\pi}{3} n' c_1' \int_0^\infty dR \frac{d\Phi_{11}(R)}{dR} \cdot \overline{F}_2'^{(11)}(R) \cdot R^3 + \frac{4\pi}{3} n' c_2' \int_0^\infty dR \cdot \frac{d\Phi_{21}(R)}{dR} \cdot \overline{F}_2'^{(12)}(R) \cdot R^3, \tag{28}$$

$$\lim_{z \to -\infty} U_2'(z) = \frac{4\pi}{3} n' c_2' \int_0^\infty dR \frac{d\Phi_{22}(R)}{dR} \cdot \overline{F}_2'^{(22)}(R) \cdot R^3 + \frac{4\pi}{3} n' c_1' \int_0^\infty dR \cdot \frac{d\Phi_{21}(R)}{dR} \cdot \overline{F}_2'^{(21)}(R) \cdot R^3, \tag{29}$$

$$\lim_{z \to +\infty} U_1'(z) = 0, \quad \lim_{z \to +\infty} U_2'(z) = 0. \tag{30}$$

Expressions for the atom work functions from the liquid mixture to the gas mixture can be obtained easily with the help of the asymptotes of the monatomic potentials for the liquid mixture and the gas mixture. Defining the atom work functions from the liquid phase to the gas phase for the atoms of the first and the second types respectively as

$$A_1 = U_1(+\infty) - U_1'(-\infty), \quad A_2 = U_2(+\infty) - U_2'(-\infty), \tag{31}$$

we obtain

$$A_1 = \frac{4\pi}{3}\int_0^\infty dR \cdot R^3 \frac{d\Phi_{11}(R)}{dR}\left[nc_1\overline{F}_2^{(11)}(R) - n'c_1'\overline{F}_2'^{(11)}(R)\right] +$$
$$+ \frac{4\pi}{3}\int_0^\infty dR \cdot R^3 \frac{d\Phi_{21}(R)}{dR}\left[nc_2\overline{F}_2^{(12)}(R) - n'c_2'\overline{F}_2'^{(12)}(R)\right], \tag{32}$$

$$A_2 = \frac{4\pi}{3}\int_0^\infty dR \cdot R^3 \frac{d\Phi_{22}(R)}{dR}\left[nc_2\overline{F}_2^{(22)}(R) - n'c_2'\overline{F}_2'^{(22)}(R)\right] +$$
$$+ \frac{4\pi}{3}\int_0^\infty dR \cdot R^3 \frac{d\Phi_{21}(R)}{dR}\left[nc_1\overline{F}_2^{(21)}(R) - n'c_1'\overline{F}_2'^{(21)}(R)\right]. \tag{33}$$

The obtained expressions for the atom fork functions of the components from the liquid mixture to the gas mixture (32), (33) can be used for a study of equilibrium conditions of two-phase two-component system and also for a study of the absolute stability criteria of mixtures.

## 3. THE CONNECTION OF THE GENERAL EQUATION OF STATE FOR LIQUIDS AND THE ATOM WORK FUNCTIONS INTO VACUUM

The general equation of state for a binary mixture of two simple liquids can be written as [6]

$$\frac{pV}{Nk_BT} = 1 - \frac{2\pi n}{3k_BT}\sum_{\alpha=1}^{2}\sum_{\beta=1}^{2}c_\alpha c_\beta \int_0^\infty \frac{d\Phi_{\alpha\beta}(R)}{dR}\overline{F}_2^{(\alpha\beta)}(R)R^3 dR. \tag{34}$$

On the other hand, using the expressions (32) and (33) for the atom work functions from the liquid mixture to vacuum we obtain

$$A_1^{l-v} = \frac{4\pi}{3}nc_1\int_0^\infty dR\frac{d\Phi_{11}(R)}{dR}\cdot\overline{F}_2^{(11)}(R)\cdot R^3 + \frac{4\pi}{3}nc_2\int_0^\infty dR\cdot\frac{d\Phi_{21}(R)}{dR}\cdot\overline{F}_2^{(12)}(R)\cdot R^3, \tag{35}$$

$$A_2^{l-v} = \frac{4\pi}{3}nc_2\int_0^\infty dR\frac{d\Phi_{22}(R)}{dR}\cdot\overline{F}_2^{(22)}(R)\cdot R^3 + \frac{4\pi}{3}nc_1\int_0^\infty dR\cdot\frac{d\Phi_{21}(R)}{dR}\cdot\overline{F}_2^{(21)}(R)\cdot R^3. \tag{36}$$

By comparing the expressions (34) and (35), (36) we can easily obtain the connection of the general equation of state for liquid mixtures and the atom work functions from the mixture to vacuum

$$p = n\left[k_BT - \frac{1}{2}\sum_{\alpha=1}^{2}c_\alpha A_\alpha^{l-v}\right]. \tag{37}$$

The equation (37) has a simple physical meaning. Accordingly to (37) all contributions to the pressure (37) have energetic senses. The first term in the right part of (37) corresponds to accounting of a kinetic energy of atoms, the second and the third terms correspond to accounting of the difference in levels of the self-consistent monatomic potential for atoms of each component in vacuum far from the surface of the mixture and the levels of monatomic potentials far from the surface in the liquid. We must notice that using the discovered connection between pressure in the mixture of simple liquids and the atom work functions from mixtures to vacuum it is possible to express all thermodynamic quantities connected with pressure in terms of the atom work functions from mixtures to vacuum or, in other words, with differences in levels of the monatomic potentials in mixtures and in vacuum.

## 4. THE ABSOLUTE STABILITY CRITERIA OF MIXTURES WITH RESPECT TO THE ATOMIZATION

For all condensed systems is typical a limited stability with respect to thermodynamic conditions in which they can exist. This fact is expressed in existing of limiting points and lines of absolute stability and in existing of metastable states [3,9]. We can consider the problem of the absolute stability criteria for mixtures in limiting points of stability of the first type [1-4] using the developed theory of the atom work functions from liquid mixtures. As was shown in the works [1,2] the stability criterion of a simple liquid in limiting points of first type corresponds to the atomization condition of the liquid. Therefore, it is naturally to make an extension of the stability criterion of a simple liquid obtained in [3,4] in case of liquid mixtures. Using the fact that states of the liquid in limiting points of the first type correspond to the atomization condition of the liquid [1,2] one can write the atomization conditions for each component of the mixture as the equality of the atom work functions and their kinetic energies

$$A_1^{l-v} = k_B T, \quad A_2^{l-v} = k_B T, \tag{38}$$

where for the atom work functions of particular components we use the expressions (35) and (36). We must notice that the relations (38) generalize the stability criteria of a simple liquid [3,4] in limiting points of the first type in case of liquid mixtures. The stability criteria (38) can hold true for each component separately. This implies that under certain thermodynamic conditions, atoms of one type can leave the mixture spontaneously. If both equalities (38) hold true, than this means that atoms of both components can leave the mixture, in other words, under such conditions the mixture is absolutely unstable with respect to the atomization of the whole mixture, but not only with respect to one particular component. Search of the atomization conditions for particular components and for the whole mixture is important for various technical and technological applications.

The atom work functions from the mixture to vacuum are functions of the concentrations, of the particle number density and of temperature. Therefore, the atomization conditions (38) can be written as

$$A_1^{l-v}(c_1, n, T) = k_B T, \quad A_2^{l-v}(c_1, n, T) = k_B T. \tag{39}$$

It is clear than, that the atomization condition for each component in the predetermined temperature can be reached on the plane $(c_1, n)$ not only in particular points but on lines. Similarly, if the particle number density is predetermined, then particular equalities (39) can be held true on the plane $(c_1, T)$ on the separate lines. Study of the position of such lines on the phase planes can compose an interesting class of problems.

Using the expressions (32) and (33) for the atom work functions from the liquid phase to the gas phase, we can formulate conditions of equality of the concentrations for one of the components in the liquid and in the gas phases as the equality of the appropriate atom work function to the thermal energy, i.e. $A_1 = k_B T$, or $A_2 = k_B T$. Realization of one of such conditions corresponds to possibility of a spontaneous equalization of the concentrations of the components in the liquid and in the gas phases due to the energy of the atoms thermal motion.

## 5. CONCLUSION

The basis of the developed microscopic theory of the atom work function from the mixture of simple liquids to the gas mixture is the equations for the unary atom distribution functions. The problem of calculation of the density profiles of atoms in the whole region $z \in (-\infty, +\infty)$ requires a self-consistent solution of the system of the equations (15). At the same time we must take into account that the pair correlation functions depend on the density profiles of atoms and on the correlation functions of next orders. Inasmuch as the explicit expressions of the pair correlation functions for inhomogeneous liquids are unknown, than a solution of the system of the equations (15) in such formulation is impossible.

The obtained stability criteria of liquid mixtures with respect to atomization of the mixture components (38) allows to state that on phase diagrams of mixtures exist limiting lines which consist of limiting points of the first type. Such lines discriminate on phase diagrams regions of absolutely unstable states with respect to atomization of particular components or with respect to atomization of the whole phase (a simultaneous realization of the conditions (39)).

Another significant result of the work concerns the obtained connection between the general equation of state for liquid mixtures and for gas mixtures and the atom work functions of mixture components. Such connection indicates the fundamental significance of levels of the self-consistent monatomic potential created by all atoms of the inhomogeneous system inside the system and in a vacuum far from system borders. The ratio of these levels of the monatomic potentials is expressed in a contribution to the general equation of state of mixture. This contribution is one-particle by its nature and it takes into account pair interactions of atoms and correlations while creating such potential. The character of activation processes in multicomponent inhomogeneous systems is strongly dependent on levels of the monatomic potentials. The widespread use of multicomponent systems in technology, biology emphasizes significance of a study of the monatomic potentials for each component of the mixture.